\documentclass[9pt,twocolumn,twoside]{opticajnl}

\journal{ol} 
\setboolean{shortarticle}{true}

\title{Efficient and robust chiral discrimination by invariant-based inverse engineering}

\author[1]{Hang Xu}
\author[1,2]{Xue-Ke Song}
\author[1]{Liu Ye}
\author[1,3]{Dong Wang}
\affil[1]{School of Physics and Optoelectronics Engineering, Anhui University, Hefei 230601, China}
\affil[2]{Corresponding author: songxk@ahu.edu.cn}
\affil[3]{Corresponding author: dwang@ahu.edu.cn}

\begin{abstract}
We propose an accurate and convenient method to achieve 100$\%$ discrimination of chiral molecules with Lewis-Riesenfeld invariant. By reversely designing the pulse scheme of handed resolution, we obtain the parameters of the three-level Hamiltonians to achieve this goal. For the same initial state,
we can completely transfer its population to one energy level for left-handed molecules, while transfer it to another energy level for right-handed molecules. Moreover, this method can be further optimized when errors exist, and it shows that the optimal method are more robust against these errors than the counterdiabatic and original invariant-based shortcut schemes. This provides an effective, accurate, and robust method to distinguish the handedness of molecules.
\end{abstract}

\begin{document}

\maketitle

Chirality, which was first proposed by Pasteur in 1848 \cite{Pasteur} originating from symmetry breaking \cite{symmetry2}, is a very important concept or attribute in natural science. It has attracted extensive attentions in specific fields of physics, materials science \cite{material}, chemistry \cite{chem}, biology \cite{bio}, and medicine \cite{medi}. In principle, when the atomic distribution and chemical bond structure of two molecules are symmetrical in the mirror image but cannot coincide, these molecules possess chirality with left ($L$) handedness or right ($R$) handedness. Generally, molecules with different chirality show the same physical and chemical properties. However, in some specific cases, they show dramatically opposite properties, especially biological activity \cite{activity1}. The drug molecules must match the geometric structure of the receptor (reactive substance) molecules in order to have the proper efficacy.

In recent years, there are many studies \cite{chiralrap,chiralcd,chiraltwolevel,chiralcp1,chiralcp2,wu1,wu2,wu3,xia1,li2,shu1} to use quantum coherent manipulation techniques to realize the effective discrimination of chiral molecules, including adiabatic passages \cite{RAP2}, counter-diabatic driving \cite{CD1,CD2,CD3,CD5},  composite pulses \cite{CP1,CP2,CP3}, etc. 
In 2019, Vitanov \emph{et al.} \cite{chiralcd} proposed an efficient chiral resolution using delayed pulses based on the principle of counter-diabatic quantum driving.
In 2019, Ye \emph{et al.} \cite{chiraltwolevel} showed two dynamic methods to achieve inner-state enantioseparation in the case that the handedness system is reduced to a effective two-level system.
In 2020, Torosov \emph{et al.} \cite{chiralcp1} introduced a method for the chiral molecule detection using sequences of three pulses, and the composite pulses are used to realize the robustness to the area error.

In this paper, we propose an efficient and robust chiral resolution method based on optimal Lewis-Riesenfeld invariant (LRI) shortcut. For the three-level Hamiltonians of the left-handed and right-handed molecules, we can design the invariants of the corresponding $L$ and $R$ systems \cite{LRI1,LRI2,LRI2.1,pertur,LRI3,LRI4}, respectively. The systems are evolved along eigenstates of their respective invariants from the same initial energy level, while they will reach to different final energy levels with regard to different chiral molecules. This means that a 100$\%$ chiral resolution is achieved. The advantage of LRI is that it has a large parameter selections to be further optimized with respect to various control errors. Taking systematic and detuning errors into account, we find that the optimal invariant shortcut scheme are more robust against these errors compared to the  counter-diabatic and the original invariant shortcuts.

\begin{figure}[ht]
  \centering
  \includegraphics[width=1\linewidth]{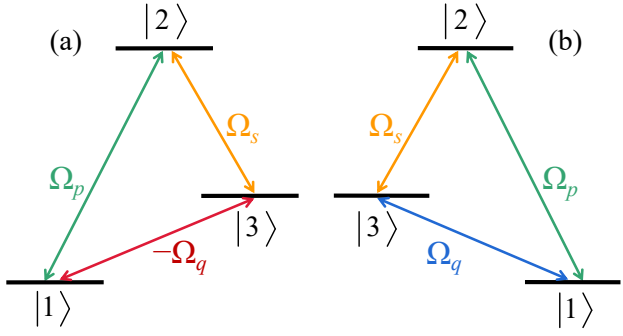}
  \caption{Schematic diagram of chiral molecules with $L$ (a) and $R$ (b) handedness in three different energy levels. Their dipole transitions are mirror symmetric, with the same ${\Omega _p}$ and ${\Omega _s}$ but the ${\Omega _q}$ with opposite sign.}
  \label{Fig1}
\end{figure}

Let us consider a typical cyclic three-level system \cite{cycli}, as shown in Fig.~\ref{Fig1}. 
The Hamiltonian, in the bases $\left\{ {\left| 1 \right\rangle ,\left| 2 \right\rangle ,\left| 3 \right\rangle } \right\}$, reads
\begin{eqnarray}
H_0^{L,R} = \hbar \left( {\begin{array}{*{20}{c}}
0&{{\Omega _p}}&{ \mp {\Omega _q}{e^{i\gamma }}}\\
{\Omega _p}&0&{{\Omega _s}}\\
{ \mp {\Omega _q}{e^{ - i\gamma }}}&{\Omega _s}&0
\end{array}} \right),
\end{eqnarray}
where the superscripts $L$ and $R$ denote the left-handedness and right-handedness. ${\Omega _p}, {\Omega _s}$ , and ${\Omega _q}$ represent the Rabi frequencies of the three energy level transitions, respectively. The sign $-$ or $+$ of ${\Omega _q}$ represents $L$ or $R$ handedness. $\gamma$ is the phase of ${\Omega _q}$, in this paper, we set $\gamma=\pi/2$ and ${\Omega _p} = {\Omega _s}=\Omega$. Therefore, the simplified Hamiltonian is
\begin{eqnarray}
{H^{L,R}} = \hbar \left( {\begin{array}{*{20}{c}}
0&\Omega &{ \mp i\Omega _q}\\
\Omega &0&\Omega \\
{ \pm i\Omega _q}&\Omega &0
\end{array}} \right),
\end{eqnarray}
In order to achieve accurate chiral resolution, the goal is that after applying the same specific pulse to the two chiral systems, the final state of the left-handedness system is completely at one energy level, and the final state of the right-handedness system is completely at another energy level, so that we can determine its chirality by measuring the energy spectrum of the system.

First, we consider the $L$ chiral system. The invariant is
\begin{eqnarray}
{I^L} = \left( {\begin{array}{*{20}{c}}
0&{\sin \varphi \sin\theta }&{ - i\cos \varphi }\\
{\sin \varphi \sin\theta }&0&{\sin \varphi \cos \theta }\\
{i\cos \varphi }&{\sin \varphi \cos \theta }&0
\end{array}} \right).
\end{eqnarray}
The eigenstates of the invariant are
\begin{eqnarray}
\begin{array}{l}
\left| {\phi _0^L} \right\rangle  = \left( {\begin{array}{*{20}{c}}
{ - \sin \varphi \cos \theta }\\
{i\cos \varphi }\\
{\sin \varphi sin\theta }
\end{array}} \right),\\
\left| {\phi _ \pm ^L} \right\rangle  = \frac{1}{{\sqrt 2 }}\left( {\begin{array}{*{20}{c}}
{\cos \varphi \cos \theta  \pm i\sin \theta }\\
{i\sin \varphi }\\
{ - \cos \varphi \sin \theta  \pm i\cos \theta }
\end{array}} \right),
\end{array}
\end{eqnarray}
with corresponding eigenvalues ${\mu _0} = 0$ and ${\mu _ \pm } =  \pm 1$. By solving the dynamical equation \cite{LRI3}, the following constraint conditions are obtained:
\begin{eqnarray}
\begin{array}{l}
\Omega  = \dot \varphi /(\sin \theta  - \cos \theta ),\\
{\Omega _q}= \dot \varphi \cot \varphi (\sin \theta  + \cos \theta )/(\sin \theta  - \cos \theta ) - \dot \theta ,
\end{array}\label{constraint}
\end{eqnarray}
where the dot represents the derivative with respect to time. When the above conditions are satisfied, we can write the general solution $\left| {\psi^L (t)} \right\rangle$ of Schr\"{o}dinger \cite{LRI1} as
\begin{eqnarray}
\left| {\psi^L (t)} \right\rangle  = \sum\limits_ {j=0,\pm}  {{B_j}{e^{i{\eta _j}(t)}}\left| {\phi _j^L}(t) \right\rangle } ,
\label{propagator}
\end{eqnarray}
where ${B_j}$ are time-independent constants, and ${\eta _j}(t)$ are the so-called LR phases which satisfy
\begin{eqnarray}
{{\dot \eta }_j}(t) = \frac{1}{\hbar }\left\langle {{\phi _j^L}(t)} \right|i\hbar \frac{\partial }{{\partial t}} - {H^{L}}\left| {{\phi _j^L}(t)} \right\rangle.\label{lrphase}
\end{eqnarray}
Thereby, we can get
\begin{eqnarray}
\begin{array}{l}
{\eta _0}(t) = 0,\\
{\eta _ \pm }(t) =  \pm \int_0^t {dt'\left[ {\dot \varphi \csc \varphi (\sin \theta  + \cos \theta )/(\cos \theta  - \sin \theta )} \right]} .
\end{array}\label{Lphase}
\end{eqnarray}
It can be seen from the Eq.~(\ref{propagator}) that if the $L$-handed system is initially in an eigenstate $\left| {\phi _j^L}(t) \right\rangle$, it will also be in this eigenstate at any time after time evolution. As for the eigenstate $\left| {\phi _0^L}(t) \right\rangle$, if the boundary conditions of the parameters are chosen as
\begin{eqnarray}
\varphi (0) = 0,\;\;\;\varphi (T) = \pi /2,\;\;\;\theta (T) = \pi /2,
\label{boundary}
\end{eqnarray}
where $T$ is final time moment, the $L$ system will completely transfer to the level $\left| 3 \right\rangle$ if initially in the level $\left| 2 \right\rangle$.
Second, let us consider the $R$ system. We set its invariant as
\begin{eqnarray}
{I^R} = \left( {\begin{array}{*{20}{c}}
0&{\sin\varphi \cos \theta }&{i\cos \varphi }\\
{\sin\varphi \cos \theta }&0&{\sin \varphi \sin\theta }\\
{ - i\cos \varphi }&{\sin \varphi \sin\theta }&0
\end{array}} \right).
\end{eqnarray}
Similarly, we can obtain the eigenstates of this invariant ${I^R}$:
\begin{eqnarray}
\begin{array}{l}
\left| {\phi _0^R} \right\rangle  = \left( {\begin{array}{*{20}{c}}
{\sin \varphi sin\theta }\\
{i\cos \varphi }\\
{ - \sin \varphi \cos \theta }
\end{array}} \right),\\
\left| {\phi _ \pm ^R} \right\rangle  = \frac{1}{{\sqrt 2 }}\left( {\begin{array}{*{20}{c}}
{ - \cos \varphi \sin \theta  \pm i\cos \theta }\\
{i\sin \varphi }\\
{\cos \varphi \cos \theta  \pm i\sin \theta }
\end{array}} \right),
\end{array}
\end{eqnarray}
with corresponding eigenvalues ${\mu _0} = 0$ and ${\mu _ \pm } =  \pm 1$. For the $R$ system, we find that the parameter constraints and LR phases of $R$ system are exactly the same as those of $L$ system, as shown in Eq.~(\ref{constraint}) and (\ref{Lphase}). This means that if we drive the $L$ or $R$  system to evolve along the eigenstate $\left| {\phi _0^L} \right\rangle$ or $\left| {\phi _0^R} \right\rangle$, we can apply the same pulse scheme by inversely solving the constraint conditions in 
Eq.~(\ref{constraint}).

\begin{figure}[tbp]
  \centering
  \includegraphics[width=1.1\linewidth]{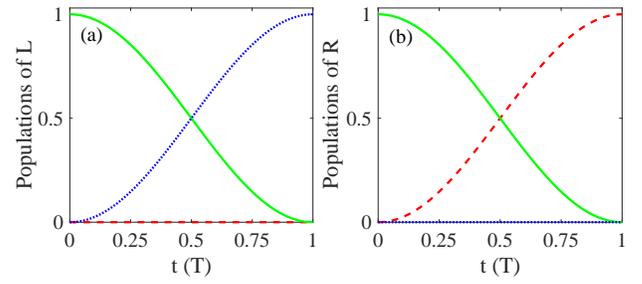}
  \caption{Schematic diagram of energy level populations of the $L (R)$ systems using SPS. (a) Populations vs the time $t$ of $L$ system; (b) Populations vs the time $t$ of $R$ system. Red dashed, green solid, blue dotted lines stand for the populations of $\left| 1 \right\rangle$, $\left| 2 \right\rangle$, and $\left| 3 \right\rangle$, respectively.}
  \label{population1}
\end{figure}

Now, we pay attention to the eigenstate $\left| {\phi _0^R}(t) \right\rangle$. If we have the same boundary condition in Eq.~(\ref{boundary}), the $R$ system will completely transfer from $\left| 2 \right\rangle$ to $\left| 1 \right\rangle$ for $t \in [0,T]$, which is completely different from the target energy level of the $L$ system. That is to say, we can apply the same pulse to a pair of $L$ and $R$ systems when they are initially at the level $\left| 2 \right\rangle$ by choosing the invariant parameters to fulfill the boundary condition in Eq.~(\ref{boundary}). This can drive the $L$ system to fully evolve to the level $\left| 3 \right\rangle$, while drive the $R$ system to fully evolve to the level $\left| 1 \right\rangle$. Finally, their handedness can be determined by measuring their energy spectrum. As a result, the 100$\%$ chiral discrimination is reached.

Here, we consider a simple parameter scheme (SPS) to show how to achieve an efficient chiral discrimination by invariant-based inverse engineering. When we choose
\begin{eqnarray}
\varphi (t) = \frac{{\pi t}}{{2T}},\;\;\;\theta (t) = \frac{\pi }{2},
\end{eqnarray}
to satisfy the boundary conditions in Eq.~(\ref{boundary}).
Inversely, we can get the parameters of Hamiltonian, from the Eq.~(\ref{constraint}), as
\begin{eqnarray}
\Omega  = \frac{\pi }{{2T}},\;\;\;{\Omega _q}= \frac{\pi }{2}\cot \frac{{\pi t}}{{2T}},
\end{eqnarray}
where $T$ is pulse duration and $t \in [0,T]$.
In Fig.~\ref{population1}, we plot the evolution curve of the level population of the $L$ and $R$ systems. It can be seen that the two systems are initially at the same level $\left| 2 \right\rangle$. At $t=T$, the population of the $L$ system completely transfers to level $\left| 3 \right\rangle$, while the population of the $R$ system completely transfers to level $\left| 1 \right\rangle$. Therefore, through measuring their energy spectrum or population of the system, we can determine its chirality: if the population of the state $\left| 3 \right\rangle$ is 1, this is a left-handed system, and if the population of the state $\left| 1 \right\rangle$ is 1, it is a right-handed system.
\begin{figure}[tbp]
  \centering
  \includegraphics[width=0.8\linewidth]{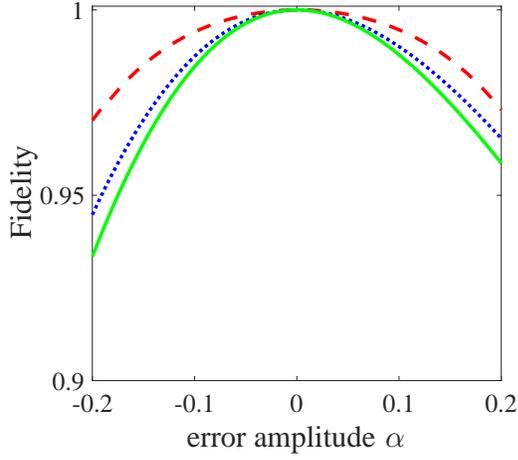}
  \caption{The systematic error amplitude $\alpha$ vs fidelity of different schemes: SPS (blue, dotted line), OSE (red, dashed line), and CD (green, solid line).}
  \label{systematic}
\end{figure}

On the other hand, when we consider the influence of control errors that may occur in the experiment on the fidelity (or discrimination) of the resolution scheme, it is necessary to optimize the LRI scheme with respect to these errors. 
We use a new Hamiltonian $H'$ to indicate the existence of errors, i.e., $H \to H' = H + {H_e}$, where ${H_e}$ is error Hamiltonian.
The fidelity is generally defined as
\begin{eqnarray}
F = {\left| {\left\langle {{\psi (T)}}
 \mathrel{\left | {\vphantom {{\psi (T)} {\psi '(T)}}}
 \right. \kern-\nulldelimiterspace}
 {{\psi '(T)}} \right\rangle } \right|^2},
\end{eqnarray}
where $\left| {\psi (T)} \right\rangle $ is target state and $\left| {\psi '(T)} \right\rangle $ is actual state of system at the final moment $T$. Using perturbation theory \cite{pertur}, we have
\begin{eqnarray}
{F^{L,R}} \approx 1 - \frac{1}{{{\hbar ^2}}}\sum\limits_ \pm  {{{\left| {\int_0^T {dt} \left\langle {\phi _0^{L,R}(t)} \right|{H_e}\left| {\phi _j^{L,R}(t)} \right\rangle {e^{i{\eta _j}(t)}}} \right|}^2}}.
\label{fidelity}
\end{eqnarray}
We first consider the influence of systematic error. 
In this case, the error Hamiltonian is described as
\begin{eqnarray}
H_e^{L,R} = \alpha {H^{L,R}},\label{error1}
\end{eqnarray}
where, $\alpha$ is a dimensionless parameter, representing the amplitude of systematic error.
Combining Eqs.~(\ref{fidelity}) and (\ref{error1}), we can get
\begin{eqnarray}
{F^L} = {F^R} = 1 - {\alpha ^2}{\left| {\int_0^T {(\dot \theta \sin \varphi  + i\dot \varphi ){e^{i{\eta _ + }(t)}}dt} } \right|^2}.
\end{eqnarray}
\begin{figure}[tbp]
  \centering
  \includegraphics[width=0.8\linewidth]{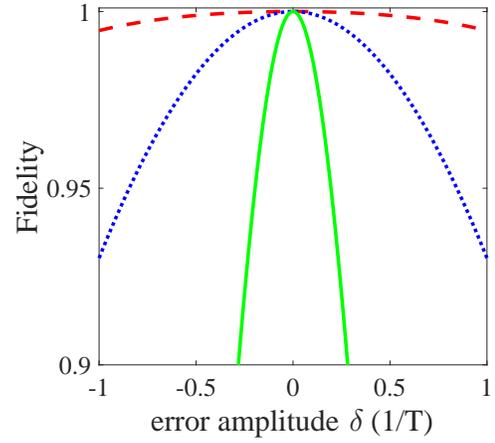}
  \caption{The detuning error amplitude $\delta$ vs fidelity of different schemes: SPS (blue, dotted line), OSD (red, dashed line), and CD (green, solid line).}
  \label{detuning}
\end{figure}
Obviously, the fidelity of the target level for the $L$ and $R$ systems is affected by the systematic error in the same way. Therefore, we only need to analyze the influence of error on the fidelity of the $L$ or $R$ system. The systematic error sensitivity is defined as
\begin{eqnarray}
{q_\alpha } =  - \frac{{{\partial ^2}{F^{L,R}}}}{{2\partial {\alpha ^2}}}{|_{\alpha  = 0}} =  - \frac{{\partial {F^{L,R}}}}{{\partial ({\alpha ^2})}}{|_{\alpha  = 0}}.
\end{eqnarray}
The smaller the sensitivity, the smaller the impact of error on fidelity. Then we have
\begin{eqnarray}
{q_\alpha } = {\left| {\int_0^T {(\dot \theta \sin \varphi  + i\dot \varphi ){e^{i{\eta _ + }(t)}}dt} } \right|^2}.
\end{eqnarray}
To meet the boundary conditions, we still choose
\begin{eqnarray}
\varphi (t) = \frac{{\pi t}}{{2T}}.
\label{psi}
\end{eqnarray}
We do not set the form of $\theta (t)$ at first, but try the Fourier series type of Ansatz with regard to the LR phase $\eta _ +$
\begin{eqnarray}
{\eta _ + }(t) =  - [n\sin(3\varphi ) - \varphi ],
\label{eta}
\end{eqnarray}
where $n$ is a real number that can be chosen freely. From the Eq.~(\ref{Lphase}), the parameter $\theta (t)$ takes the form
\begin{eqnarray}
\theta (t) = arccot\frac{{3n\cos(3\varphi )\sin\varphi  - \sin \varphi  + 1}}{{3n\cos(3\varphi )\sin\varphi  - \sin \varphi  - 1}},\label{theta}
\end{eqnarray}
which  satisfies the boundary condition $\theta (T) = \pi /2$. Based on the above equations, we can calculate the systematic error sensitivity ${q_\alpha }$ numerically. When $n=1.07$, the  systematic error sensitivity reaches the minimum value of 0.52, which is defined as the optimal scheme for systematic error sensitivity (OSS). In Fig.~\ref{systematic}, we compare the influence of systematic error on the fidelity or discrimination with several coherent control schemes, including OSS, SPS, and the counter-dabatic (CD) shortcut method in Ref.~\cite{chiralcd}. We can observe that all these schemes can achieve 100$\%$ discrimination in the absence of the error, and the  OSE scheme is the most robust against systematic error, followed by SPS, and finally CD.

Another important error in experiment is the detuning error. In this case, the error Hamiltonian is
\begin{eqnarray}
H_e^{L,R}  = \delta \hbar (\left| 3 \right\rangle \left\langle 3 \right| - \left| 1 \right\rangle \left\langle 1 \right|),
\end{eqnarray}
where $\delta$ represents the detuning amplitude, and its unit is $1/T$.
In the same way, we can obtain the fidelity as
\begin{figure}[tbp]
  \centering
  \includegraphics[width=0.8\linewidth]{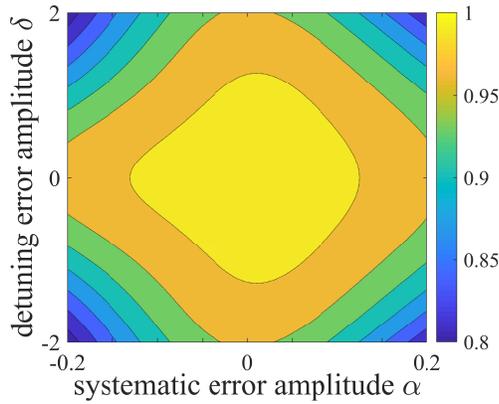}
  \caption{Fidelity $F^{L,R}$ vs the systematic error amplitude $\alpha$ and detuning error amplitude $\delta$ by LRI scheme of $n=1.10$. The yellow area in the middle corresponds to $F^{L,R} \ge 0.99$.}
  \label{fig5}
\end{figure}
\begin{eqnarray}
\begin{array}{l}
{F^L} = {F^R}\\
 = 1 - \frac{{{\delta ^2}}}{4}{\left| {\int_0^T {[\cos(2\theta )\sin(2\varphi ) + 2i\sin (2\theta )\sin\varphi ]{e^{i{\eta _ + }(t)}}dt} } \right|^2}.
\end{array}
\end{eqnarray}
And we have
\begin{eqnarray}
{q_\delta } = {\left| {\int_0^T {[\cos(2\theta )\sin(2\varphi ) + 2i\sin (2\theta )\sin\varphi ]{e^{i{\eta _ + }(t)}}dt} } \right|^2}.
\end{eqnarray}
Here, the parameters $\varphi$ and $\eta _ +$ are chosen as the same forms of Eqs.~(\ref{psi}) and (\ref{eta}). We can find that, the detuning error sensitivity reaches the minimum value 0 when $n=1.12$. We call the corresponding parameter scheme  as the optimal scheme for detuning error (OSD). In Fig.~\ref{detuning}, we compare the influence of detuning error on the fidelity or discrimination with OSD, SPS, and CD control schemes. Again, the OSD scheme is the most robust against the detuning error. Furthermore, we plot how the fidelity is affected by the systematic error and detuning error in Fig.~\ref{fig5}. It can be seen that the optimal scheme shows high robustness against these two errors with a broad range of high efficiencies over 99\% .

In conclusion, we propose a highly efficient and robust chiral discrimination method for the cyclic three-level systems of chiral molecules based on the invariant-based inverse engineering. Through applying to the same pulse on the three-level system, molecules with different chirality will transit to different energy levels. The $L$ system stay in $\left| 3 \right\rangle$ and $R$ system stay in $\left| 1 \right\rangle$ at the final time from the same initial state. We can realize the 100$\%$ chiral discrimination of molecules by measuring population or energy spectrum. Moreover, we can design the corresponding optimization schemes with respect to different experimental errors. By comparison, the optimization schemes are superior to the SPS and the CD schemes.

\begin{backmatter}
\bmsection{Funding} This study was supported by the National Natural Science Foundation of China (Grant No. 12004006, No. 12075001, and No. 12175001), Anhui Provincial Key Research and Development Plan (Grant No. 2022b13020004), and the Anhui Provincial Natural Science Foundation (Grant No. 2008085QA43).

\bmsection{Disclosures} The authors declare no conflicts of interest.

\bmsection{Data Availability Statement}  Data underlying the results presented in this Letter are not publicly available at this time but may be obtained from the authors upon reasonable request.

\end{backmatter}

\bibliography{sample}

\bibliographyfullrefs{sample}
\end{document}